\documentclass[reprint, superscriptaddress, amsmath, amssymb, aps, prx, floatfix]{revtex4-2}

\usepackage{braket}
\usepackage{printlen} 
\usepackage{amsthm}
\usepackage{isotope}
\usepackage[
    per-mode=symbol,
    retain-zero-uncertainty=true
    ]{siunitx}
\DeclareSIUnit{\gauss}{G}
\DeclareSIUnit{\phonons}{phonons}
\usepackage{braket}
\usepackage{color}
\usepackage{graphicx}
\graphicspath{{supp_mat_figures/}}
\usepackage{dcolumn}
\usepackage{bm}
\usepackage{hyperref}
\hypersetup{
pdftitle={Tweezer-assisted subwavelength positioning of atomic arrays in an optical cavity},
pdfauthor={QGate Team},
pdfkeywords = {Optical Tweezers, cold atoms, cavity QED},
colorlinks=true, linkcolor=[RGB]{0,0,0}, citecolor=magenta, filecolor=[RGB]{0,0,0}, urlcolor=[RGB]{0,0,0}}

\usepackage[
    per-mode=symbol,
    retain-zero-uncertainty=true
    ]{siunitx}

\usepackage{bm}
\usepackage{enumerate}
\usepackage{tikz}
\usepackage{dsfont}
\usetikzlibrary{quantikz}
\usepackage{lipsum} 
\usepackage{subcaption}
\usepackage{ragged2e}
\DeclareCaptionJustification{justified}{\justifying}
\usepackage{printlen}
\usepackage{etoolbox}
\usepackage{soul}
\makeatletter
\def\maketitle{
\@author@finish
\title@column\titleblock@produce
\suppressfloats[t]}
\makeatother
\newcounter{PRLsections}

\newcounter{PRLsubsections}[section]

\usepackage{xpatch}
\xpretocmd{\section}{\setcounter{PRLsubsections}{0}}{}{}

\usepackage{wasysym}
\usepackage[OT2,T1]{fontenc}
\DeclareSymbolFont{cyrletters}{OT2}{wncyr}{m}{n}
\DeclareMathSymbol{\Sha}{\mathalpha}{cyrletters}{"58}

\usepackage[innercaption]{sidecap}
\sidecaptionvpos{figure}{c}
\DeclareSymbolFont{yhlargesymbols}{OMX}{yhex}{m}{n} 
\DeclareMathAccent{\widehat}{\mathord}{yhlargesymbols}{"62}
\usepackage{multirow}
\usepackage{array}
\newcolumntype{L}[1]{>{\raggedright\let\newline\\\arraybackslash\hspace{0pt}}m{#1}}
\newcolumntype{C}[1]{>{\centering\let\newline\\\arraybackslash\hspace{0pt}}m{#1}}
\newcolumntype{R}[1]{>{\raggedleft\let\newline\\\arraybackslash\hspace{0pt}}m{#1}}

\usepackage{pgfplots}
\usepackage{pgf}

\newcommand{\be}{\begin{equation}} 
\newcommand{\ee}{\end{equation}}

\newtoggle{arXiv} 

\toggletrue{arXiv}      
\begin{document}
\preprint{APS/123-QED}

\title{Tweezer-assisted subwavelength positioning of atomic arrays in an optical cavity}

\author{Matthias Seubert}
\email[Contact author: ]{matthias.seubert@mpq.mpg.de}
\affiliation{Max-Planck-Institute of Quantum Optics, Hans-Kopfermann-Strasse 1, 85748 Garching, Germany}
\author{Lukas Hartung}
\affiliation{Max-Planck-Institute of Quantum Optics, Hans-Kopfermann-Strasse 1, 85748 Garching, Germany}
\author{Stephan Welte}
\altaffiliation{Present address: 5. Physikalisches Institut, Universität Stuttgart, Pfaffenwaldring 57, 70569 Stuttgart, Germany}
\affiliation{Max-Planck-Institute of Quantum Optics, Hans-Kopfermann-Strasse 1, 85748 Garching, Germany}
\affiliation{Institute for Quantum Electronics, ETH Z\"urich, Otto-Stern-Weg 1, 8093 Zürich, Switzerland}
\author{Gerhard Rempe}
\affiliation{Max-Planck-Institute of Quantum Optics, Hans-Kopfermann-Strasse 1, 85748 Garching, Germany}
\author{Emanuele Distante$^{1,}$}
\email[Contact author: ]{emanuele.distante@icfo.eu}
\affiliation{ICFO-Institut de Ciencies Fotoniques, The Barcelona Institute of Science and Technology, 08860 Castelldefels, Barcelona, Spain}

 
\begin{abstract}
\medskip \noindent

We present an experimental technique that enables the preparation of defect-free arrays of \isotope[87]{Rb} atoms within a microscopic high-finesse optical standing-wave cavity. By employing optical tweezers, we demonstrate atom positioning with a precision well below the cavity wavelength, a crucial requirement for cavity-QED experiments in which maximum atom-cavity coupling strength is required. We leverage our control capabilities to assemble an array of up to seven atoms with an efficiency that exceeds previous probabilistic methods by 4 orders of magnitude. The atoms are subsequently transferred from the tweezer array to a two-dimensional intracavity optical lattice that offers enhanced coherence for spin qubits while maintaining strong atom confinement. Our system overcomes the efficiency limitations of previous probabilistic loading techniques of cavity-coupled atom arrays and opens the path to multiqubit quantum networks with atoms strongly coupled to optical cavities.   
\end{abstract}

\maketitle 

\section{Introduction} 

Driven by a wide range of applications ranging from quantum communication \cite{lo2014} to distributed computing \cite{buhrman2003, Jiang2007, monroe2014} and sensing \cite{gottesman2012, komar2014}, the practical implementation of quantum networks has now challenged scientists for over two decades. The envisioned architectures rely on using flying photons to share entangled states among distant network nodes, where stationary memory qubits can store and process quantum information \cite{kimble2008,sangouard2011,reiserer2015,wehner2018}. A major challenge consists of distributing high-quality entanglement on time scales faster than the typical decoherence time of the memory qubits. Unfortunately, unavoidable optical losses within the nodes and along the network links, in combination with the nonvanishing communication time, increase the entanglement distribution time dramatically, while different noise sources, such as light polarization or phase fluctuations in the optical link, degrade the quality of the distributed entanglement, thereby limiting the utility of the networks. While early demonstrations have primarily focused on networks with one \cite{moehring2007,ritter2012,hofmann2012,bernien2013,delteil2016} or two stationary qubits per node \cite{kalb2017}, a possible solution consists of developing nodes comprising many controllable qubits \cite{covey2023}. These advanced network processing nodes can implement multiplexed quantum communication \cite{sangouard2011, krutyanskiy2024, hartung2024} in order to speed up the entanglement distribution time and, furthermore, to implement error correction \cite{roffe2019} and entanglement distillation \cite{kalb2017}. 

Several promising platforms are being developed for multiqubit network nodes, such as trapped ions \cite{krutyanskiy2024, canteri2024} or color centers \cite{knaut2024} and dopants \cite{Chen2020} in solid-state host materials. Among these, neutral atoms coupled to high-finesse optical cavities stand out for their combination of strong light-matter coupling \cite{reiserer2015}, long coherence times \cite{korber2018}, and fast and efficient light-matter-entanglement generation \cite{thomas2022}. These features have led to the demonstration of several quantum network primitives, including quantum teleportation \cite{langenfeld2021a}, repeaters \cite{langenfeld2021}, and distributed quantum gates \cite{daiss2021}. To realize these applications, single atoms must be loaded and positioned within the cavity mode with subwavelength precision to maximize the atom-cavity coupling strength. For up to one or two atoms, this can be accomplished using probabilistic loading schemes \cite{nussmann2005}. However, these methods become exponentially inefficient as the number of atoms increases. Reconfigurable optical tweezers \cite{kim2016, barredo2016, endres2016} offer a promising solution, as they enable precise control over individual atom positions. This technology has transformed neutral atom-based quantum computing \cite{henriet2020} by facilitating the creation of defect-free atomic arrays with thousands of atoms \cite{manetsch2024}. Integrating it with cavity-based systems would mark a significant advancement in scaling neutral-atom quantum network nodes \cite{dhordjevic2021, deist2022_2, yan2023, liu2023, hartung2024}.

In this paper, we present our experimental setup that merges neutral atoms in optical tweezers with a high-finesse cavity for preparing intracavity defect-free arrays of atoms. We demonstrate subwavelength atom positioning and show an improvement by 4 orders of magnitude in atom-loading efficiency compared to previously employed probabilistic loading schemes of cavity-coupled atom arrays. We showcase this improvement for arrays containing up to seven atoms. Finally, we demonstrate a controlled transfer of all atoms to a two-dimensional (2D) optical lattice trap \cite{young2022, norcia2024}. Compared to approaches employing only optical tweezers \cite{dhordjevic2021,deist2022_2,liu2023}, the lattice reduces major sources of decoherence such as light scattering and tweezer-induced virtual magnetic fields \cite{rosenfeld2008,thompson2013}, improves the long-term stability of the atom positioning relative to the cavity field, and allows for strong confinement at lower laser powers. Utilizing the strengths of the tweezers and the optical lattice traps, our system paves the way for multiqubit quantum network nodes based on atom arrays strongly coupled to optical cavities.

The paper is structured as follows: Sec. \hyperref[sec:Experimental_setup]{II} describes our experimental setup, Sec. \hyperref[sec:Subwavelength_positioning]{III} demonstrates the subwavelength positioning of a single atom in the cavity mode, and Sec. \hyperref[sec:Assembling_ordered]{IV} describes the protocol for assembling ordered atom arrays and the controlled transfer of the atoms to the optical lattice. We conclude with a discussion of the implications and prospects of our work in Sec. \hyperref[sec:Conclusion]{V}. Details of our optical tweezers are presented in the \hyperref[sec:Appendix]{Appendix}. 

\section{Experimental setup}
\label{sec:Experimental_setup}

\begin{figure}[t]
    \centering
    \includegraphics[width=1.0\columnwidth]{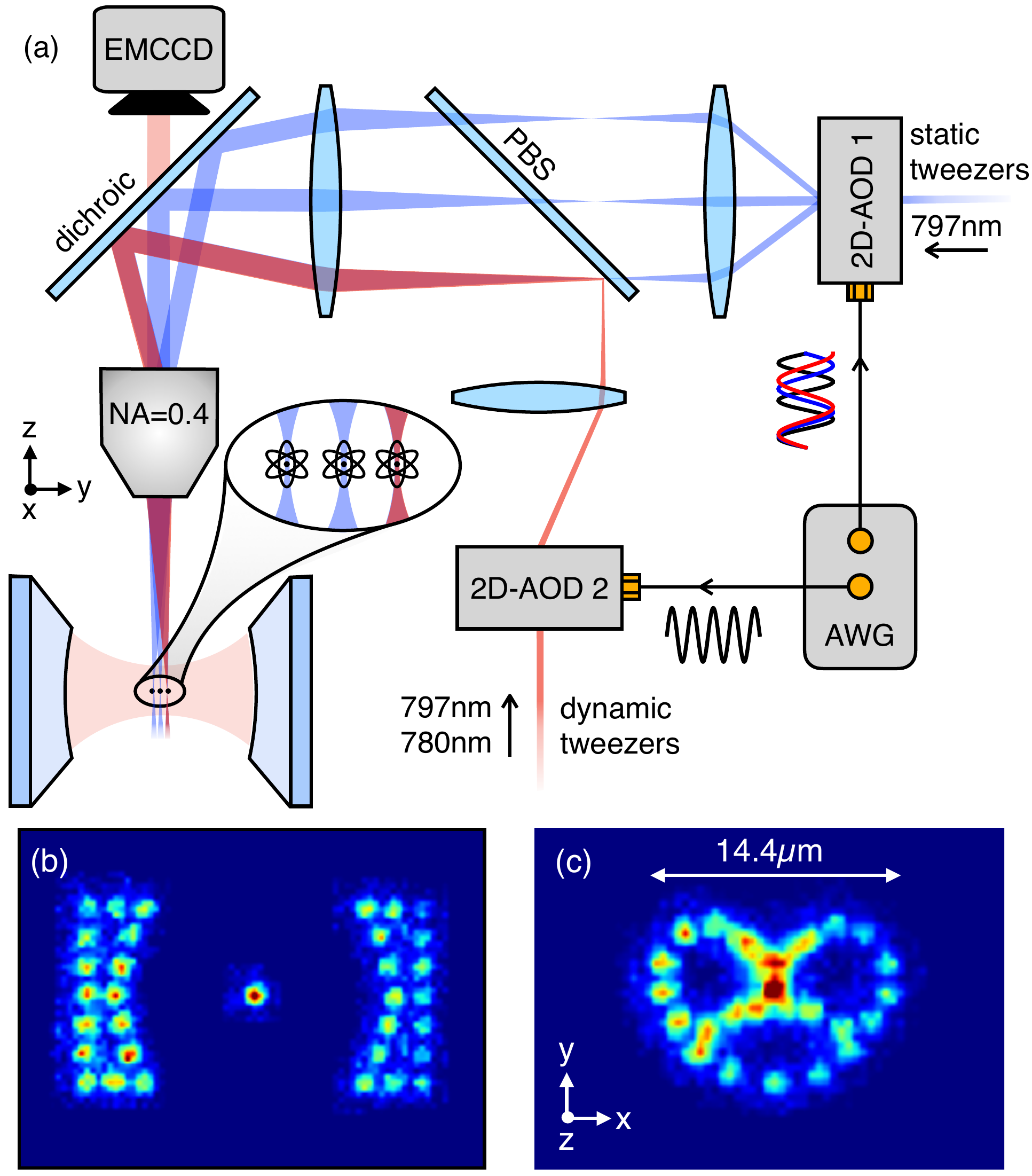}
    \caption{(a) The experimental setup. We employ two two-dimensional (2D) acousto-optic deflectors (AODs) to prepare \textit{static} and \textit{dynamic} tweezers for trapping and rearranging atoms within the cavity mode. The AODs are supplied with radio-frequency (rf) signals from an arbitrary-waveform generator (AWG). The atoms are imaged with an electron-multiplying charge-coupled device (EMCCD) camera and we employ feedback from the camera picture analysis to the AWG supplying the AODs to adjust the tweezer positions. (b),(c) The generation of two example geometries of atom arrangements in two dimensions within the cavity mode. Both pictures are the sum of several pictures each showing one single atom at different locations.}
    \label{fig:setup}
\end{figure}

In Fig.~\ref{fig:setup}, we present our experimental setup that combines the neutral-atom optical-tweezer technology with a high-finesse optical cavity. The cavity is a  Fabry-P\'{e}rot resonator made of two tapered mirrors with transmissivities $\qty{4}{ppm}$ (parts per million) and $\qty{92}{ppm}$, with $\qty{7}{ppm}$ round-trip losses, resulting in a finesse of $\mathcal{F}=61000(2000)$. It has a length of $\qty{485}{\micro \meter}$  and a mode waist of $\omega_{\text{C}}= \qty{29}{\micro \meter}$. The cavity-QED parameters are $\left\{ g_0, \kappa, \kappa_{\text{out}},\gamma \right \} = 2 \pi \times \left\{\qty{7.8}{},\qty{2.5}{}, \qty{2.3}{},\qty{3}{}\right\}\qty{}{\mega \Hz}$, where $g_0$ is the atom-cavity-field interaction strength given for the $\ket{5S_{1/2}, F=2,m_\text{F}=2}\rightarrow \ket{5P_{3/2}, F=3,m_\text{F}=3}$ transition, $\kappa$ and $\kappa_\text{out}$ are the total cavity-field decay rate and  the decay rate through the outcoupling mirror, and $\gamma$ is the atomic polarization decay rate.

The optical tweezers are formed by tightly focusing a laser onto the cavity plane using an objective with numerical aperture NA=0.42. The presence of the cavity mirrors limits it to NA=0.31, resulting in a diffraction-limited waist of $\qty{1.08}{\micro \meter}$. Depending on the experiment, we select the wavelength of the tweezer beams in the range of \qtyrange{797}{807}{\nano \meter}. We generate two sets of tweezers, referred to as \textit{dynamic} and \textit{static} tweezers. Each tweezer is controlled independently using two 2D acousto-optic deflectors (AODs). The laser beams from both tweezers are overlapped on a polarizing beam splitter (PBS). Both tweezer paths include a magnifying telescope in a $4f$ configuration to enlarge the laser-beam diameter (magnification M=8) and image the AOD aperture onto the back aperture of the objective. Here, we only summarize the main specifications of our tweezers, while giving further details in the Appendix: the measured waists in the \textit{x-y} plane are $\{ w_\text{x}, w_\text{y} \} = \{\qty{1.28(11)}{\micro \metre}, \, \qty{1.49(13)}{\micro \metre}\}$ (see Appendix \hyperref[subsec:Super-resolving]{2}), larger than the diffraction limit but compatible with the one estimated using the initial laser spot size, the specified magnifying telescope, and the effective NA of the objective. For a selected wavelength of \qty{797}{\nano \meter}, this results in a trap depth of $U(P_\text{T})/k_\text{B} = 1.76 \,\mathrm{mK} \times \,P_\text{T}/\mathrm{mW}$, where $P_\text{T}$ is the individual tweezer-trap power and $k_\text{B}$ is the Boltzmann constant. 

Our setup includes an electron-multiplying charge-coupled device (EMCCD) to image the atoms by collecting atomic fluorescence via the objective. To separate the fluorescence from the tweezer light, we use a custom-made bandpass mirror that transmits \qty{70(2)}{\percent} of the light at \qty{780}{\nano \meter} and reflects nearly $\qty{100}{\percent}$ in the near-infrared band \qtyrange{770}{810}{\nano \meter}. The coordinate system is defined by the cavity axis (y) and the axis of the tweezers (z). The optical tweezers provide precise control over the position of individual atoms within the cavity mode (the x-y plane), enabling the creation of customized atom arrangements. To showcase this, we steer a single atom to various locations and superimpose the different images to form distinct geometries, such as a configuration representing a cavity containing a single atom, or a pretzel, as shown in Figs.~\hyperref[fig:setup]{1(b)} and \hyperref[fig:setup]{1(c)}. The atom positioning relative to the cavity field is described in detail in Sec. \hyperref[sec:Subwavelength_positioning]{III}. 

\section{Subwavelength positioning of a single atom within the cavity mode}
\label{sec:Subwavelength_positioning}

\begin{figure}[t]
    \centering
    \includegraphics[width=1.0\columnwidth]{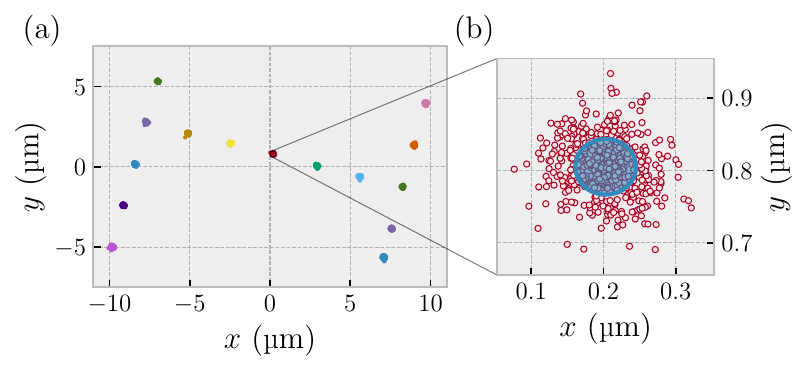}
    \caption{The positioning of atoms in the cavity mode. The atoms are positioned by applying a pair of radio-frequencies to the 2D AOD. (a) Atoms are positioned in 15 locations within the cavity mode, each color corresponding to a different pair of radio-frequencies applied to our 2D AOD used for atom positioning. (b) An enlargement in one location. Each data point is the result of a 2D Gaussian fit to the picture showing the atomic fluorescence. The standard deviation of this distribution (blue ellipse) is on average $\overline{\sigma}_\text{x} = \qty{47(5)}{\nano \metre}$ and $\overline{\sigma}_\text{y} = \qty{46(6)}{\nano \metre}$ and it represents an upper bound of the positioning error.}
    \label{fig:positioning}
\end{figure}

In this section, we demonstrate the ability to position a single atom within the cavity with a precision well below the cavity-field wavelength ($\lambda_\text{C} =\qty{780}{\nano \meter}$). Since, along the cavity axis $y$, the atom-cavity interaction strength is spatially modulated as $g(y)=g_0 \, \mathrm{sin}(2\pi y/\lambda_\text{C})$, this precision allows us to tune the cavity-atom interaction strength and phase for each atom in a loaded-atom array. To estimate the positioning precision, we repeatedly place a single atom at several predetermined locations within the cavity mode. We determine the position of the atom at each site by fitting the atomic fluorescence detected by the EMCCD camera with a 2D Gaussian distribution. Our results in Fig.~\ref{fig:positioning} indicate an average standard deviation of the atomic position per site of $\overline{\sigma}_\text{x} = \qty{47(5)}{\nano\meter}$ and $\overline{\sigma}_\text{y} = \qty{46(6)}{\nano\meter}$. Note that $\overline{\sigma}_\text{x,y}$ include the atom-position detection error, making them an upper bound of the positioning error.

\begin{figure}[t]
    \centering
    \includegraphics[width=1.0\columnwidth]{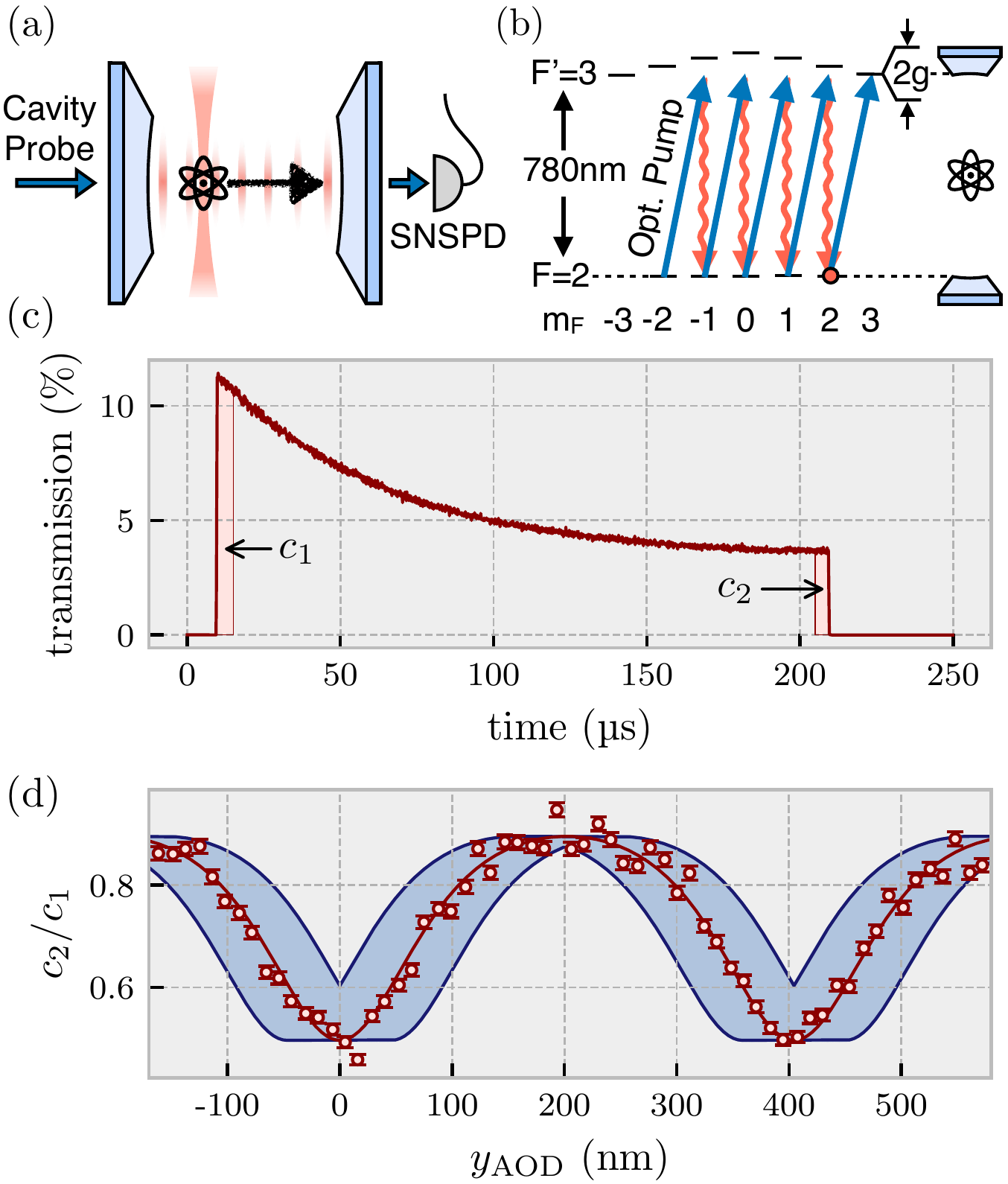}
    \caption{The transport of a single tweezer trapped atom along the cavity symmetry axis. (a) The cavity transmission is monitored with a superconducting nanowire single-photon detector (SNSPD) as a function of the atom position.
    (b) The atomic-level scheme, showing the normal-mode splitting of the coupled excited state and the probe-induced optical pumping dynamic. (c) The transmission of the cavity as a function of the optical pumping time. An atom fully pumped into the $\ket{5S_{1/2}, F{=}2,m_\text{F}{=}2}$ state suppresses the cavity transmission due to the strong atom-cavity coupling. We measure the counts $c_1$ ($c_2$) within an interval of \qty{5}{\micro \second} at the beginning (end) of the pumping procedure. 
    (d) The transmission as a function of the atom position along the $y_\text{AOD}$ axis. The $y_\text{AOD}$ axis is tilted by a small angle $\alpha = \ang{14.60(5)}$ from the cavity axis. We observe a modulation of the cavity transmission with a periodicity of $\qty{404(2)}{\nano \metre}$, compatible with an angle of $\alpha = \qty{15(1)}{\degree}$. The blue shaded area is given by  $\overline{\sigma}_\text{y} = \qty{46(6)}{\nano \metre}$ (see Fig.~\ref{fig:positioning}), which includes the atom-detection precision.}
    \label{fig:pumpingAlongCavity}
\end{figure}

As $\overline{\sigma}_\text{x,y} \ll \lambda_\text{C}$, we can showcase the possibility of tuning the effective atom-cavity interaction strength for a single atom in the cavity. To this end, we perform the experiment schematically shown in Fig.~\ref{fig:pumpingAlongCavity}. We trap an atom using an optical tweezer at \qty{807}{\nano \meter}. At this wavelength, the excited state $\ket{5P_{3/2}, F^\prime{=}3}$ experiences a large tensor light shift $\Delta_{\text{m}_\text{F}} \approx -31.6 \,\mathrm{kHz} \times m_\text{F}^2 I_\text{T}/(\mathrm{mW} / \mathrm{cm^{2}})$, with tweezer intensity $I_\text{T}$. Tuning $I_\text{T}$ allows us to set the cavity on resonance only with the $\ket{5S_{1/2}, F{=}2,m_\text{F}{=}\pm2} \rightarrow \ket{5P_{3/2}, F^\prime{=}3,m_\text{F}{=}\pm3}$ atomic transitions. Using a superconducting nanowire single-photon detector (SNSPD), we monitor the transmission of a circularly ($\sigma^+$) polarized cavity-resonant probe beam through the cavity. Simultaneously, we illuminate the atom with a repumper beam to continuously pump the atomic population into the $\ket{5S_{1/2}, F{=}2}$ state. At the beginning of the probe interval, the atom population is randomly distributed over the five $\ket{5S_{1/2}, F{=}2, m_\text{F}}$ states, and the transmission is expected to be high. During the probe interval, on average, a scattering of a $\sigma^+$ polarized probe photon transfers the atomic population toward a state with higher magnetic quantum number, eventually optically pumping the atom into the $\ket{5S_{1/2}, F{=}2, m_\text{F}{=}2}$ state [see Fig.~\hyperref[fig:pumpingAlongCavity]{3(b)}]. When pumped into this state, the atom-cavity coupling results in a normal mode splitting of the excited state, causing a drop of the probe transmission to a value $T$($g(y)$$)=4 (\kappa-\kappa_\text{out}) \kappa_\text{out} \gamma^2/(\gamma\kappa+ g^2(y))^2$, provided that the atom is located at a position at which it couples to the cavity, i.e., $g^2(y)/(\kappa\gamma)\gg1$. In Fig.~\hyperref[fig:pumpingAlongCavity]{3(c)}, we show the probe pulse transmission as a function of time when the atom is in a  maximally coupled position. We observe that the counts $c_1$ within the \qty{5}{\micro \second} interval at the beginning of the pulse exceed those at the end of the pulse $c_2$. Note that the ratio $c_2/c_1$ depends on the random atomic state population at the beginning of the probe interval, and on the efficiency of the probe optical pumping to $\ket{5S_{1/2}, F{=}2, m_\text{F}{=}2}$. If the atom is moved along the cavity axis, the transmission is modulated with a period given by $\lambda_\text{C}/2 = \qty{390}{\nano \meter}$. In Fig.~\hyperref[fig:pumpingAlongCavity]{3(d)}, we scan the atom position along the vertical axis of the AOD, $y_\text{AOD}$, which has a small angle $\alpha = \ang{14.60(5)}$  relative to the cavity axis (see Appendix \hyperref[subsec:Super-resolving]{2}). We fit our data with $A \, T(g(y-y_0))) + c$, where $g(y) = g_\text{eff} \, \mathrm{sin}(2\pi y/\lambda_\text{C})$. Here, $A$, $c$, $g_\text{eff}$ and $y_0$ are free-fitting parameters that account for the optical pumping efficiency ($A$, $c$), the scan initial position ($y_0$), and the effective cavity coupling ($g_\text{eff}$). We estimate a modulation period of $\qty{404(2)}{\nano \meter}$, which would correspond to an angle $\alpha = \ang{15(1)}$, compatible with the expected value. These results show the possibility of precisely tuning the coupling strength $g$ for a single atom in the cavity. Combined with the ability to move atoms on time scales faster than their coherence time \cite{bluvstein2022}, this could be leveraged to realize protocols in which $g(y)$ is dynamically set for each atom in a larger array.

\section{Assembling ordered arrays of atoms}
\label{sec:Assembling_ordered}

\begin{figure}[t!]
    \centering
    \includegraphics[width=1.0\columnwidth]{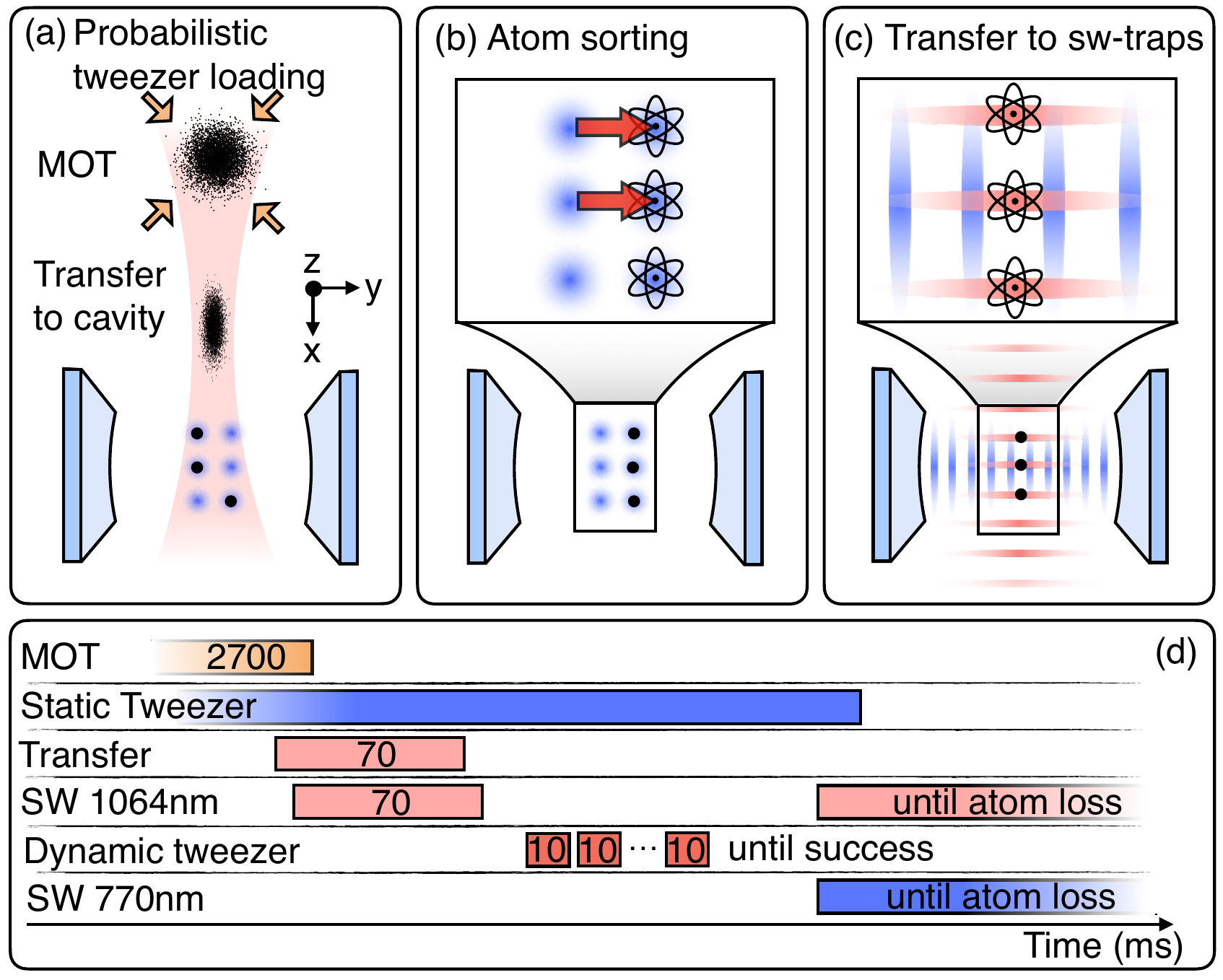}
    \caption{A schematic of the experimental protocol. (a) A cold atomic cloud is loaded into a magneto-optical trap (MOT) and transferred into the cavity mode via a transfer dipole-trap laser (red). The atoms are then probabilistically loaded into the tweezer-trap array (blue laser beams). (b) Atom sorting with the \textit{dynamic} tweezer (red) to prepare an ordered array. (c) The transfer of the ordered array into two orthogonal standing-wave (SW) traps at \qty{1064}{\nano\metre} (red) and \qty{771}{\nano\metre} (blue). (d) The timing of the tweezer loading and rearranging protocol. All times are given in units of milliseconds.   
    }
    \label{fig:fig1}
\end{figure}

To trap ordered arrays of single $\isotope[87]{Rb}$ atoms in the cavity mode, we employ a three-step protocol schematically shown in Fig.~\ref{fig:fig1}. The three steps are (\hyperref[subsec:Stochastic_loading]{A}) stochastic loading of an intracavity tweezer array [Fig.~\hyperref[fig:fig1]{4(a)}], (\hyperref[subsec:Atom_sorting]{B}) atomic sorting to a defect-free array [Fig.~\hyperref[fig:fig1]{4(b)}], and (\hyperref[subsec:Controlled_transfers]{C}) controlled transfer of the atomic array to a 2D optical lattice [Fig.~\hyperref[fig:fig1]{4(c)}]. The sequence employed is shown in Fig.~\hyperref[fig:fig1]{4(d)}. In the following, we will detail these three steps and report our results.

\subsection{Stochastic loading of an intracavity tweezer array}
\label{subsec:Stochastic_loading}

Due to the limited optical access within the cavity mode, we load atoms in a magneto-optical trap (MOT) formed \qty{14}{\milli \meter} away from the cavity center and then transfer the atoms to the cavity region. The transfer lasts  \qty{70}{\milli \second} and uses a combination of a transfer running-wave dipole trap and a standing-wave dipole trap, both operating at \qty{1064}{\nano \meter}. The transfer trap is focused between the MOT position and the cavity center, while the standing-wave trap is focused at the cavity center to a waist of \qty{12.88}{\micro \meter}. This method has proven to be an efficient and reliable approach for loading and cooling atoms within the cavity, in a region defined by the spatial size of the standing-wave trap \cite{nussmann2005}.

During the transfer, a 2D array of tweezer traps is kept constantly on by driving the \textit{static} tweezer 2D AOD with a multitone radio-frequency (rf) signal, which is optimized for a small peak-to-peak amplitude \cite{schroeder1970}. Each tweezer has a trap depth of $U_{\textit{static}}/k_\text{B}=\qty{3.5}{\milli \kelvin}$. The standing-wave and transfer traps are then ramped down and atoms are probabilistically captured by the tweezer traps. To increase the loading probability, the tweezer grid is carefully aligned with the standing-wave-trap axis to ensure maximal overlap between the atoms and the tweezer region. Throughout the loading process, the atoms are continuously cooled by polarization-gradient cooling using an optical access at a $45^{\circ}$ angle in the x-z plane. The scattered cooling light from the atoms is collected by the objective, enabling the detection of filled traps within \qty{350}{\milli \second} using the EMCCD camera.

\subsection{Atom sorting}
\label{subsec:Atom_sorting}

\begin{figure}[t]
    \centering
    \includegraphics[width=1.0\columnwidth]{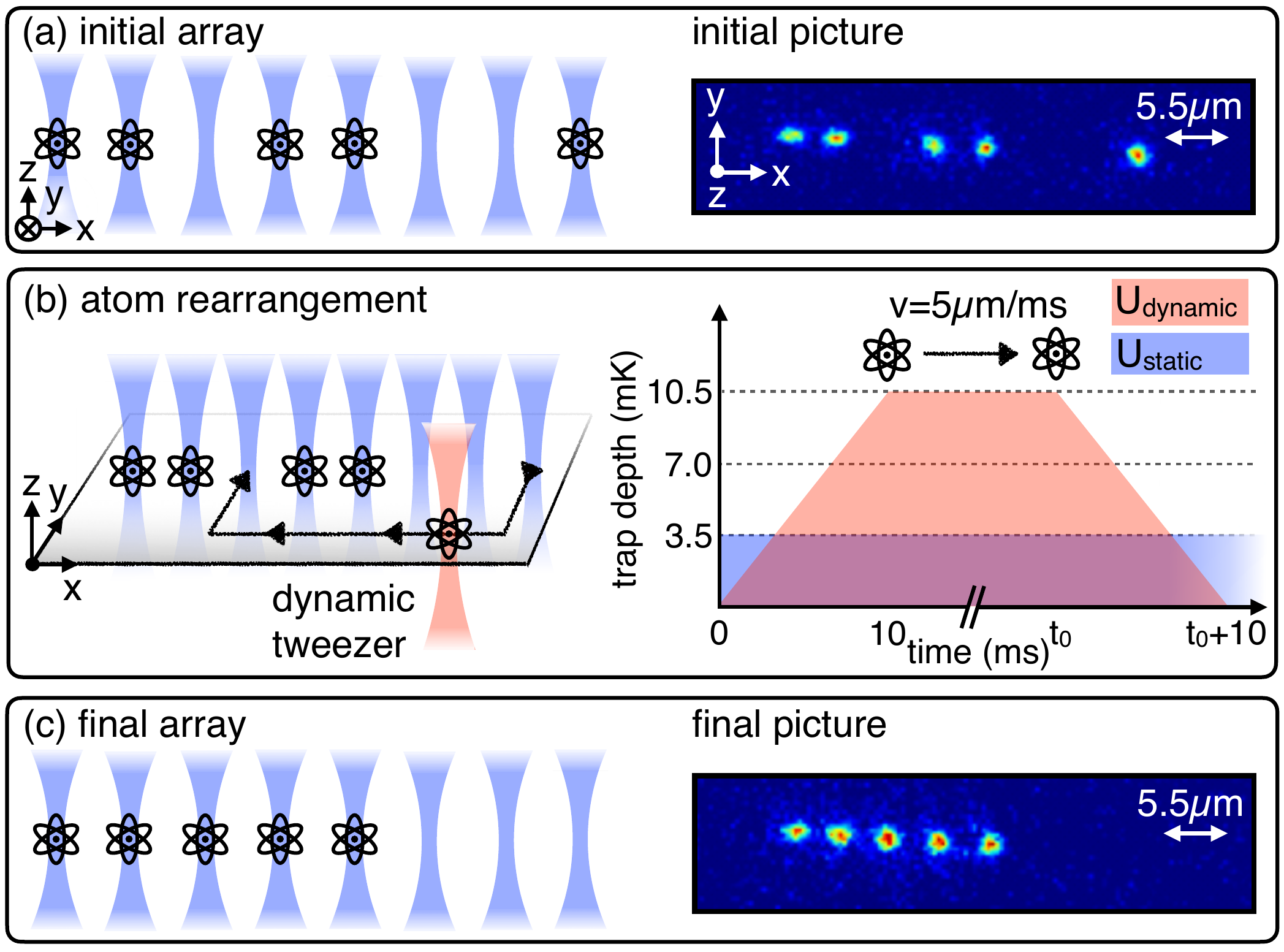}
    \caption{The rearrangement of intracavity atoms with optical tweezers. After (a) taking an initial picture and identifying the occupied tweezer traps, we (b) employ a \textit{dynamic} transport tweezer, which allows us to (c) generate an ordered array of atoms. The trap depth of the \textit{dynamic} tweezer is 3 times higher than the trap depth of the \textit{static} tweezers.  
    }
    \label{fig:fig3}
\end{figure}

\begin{figure}[t]
    \centering
    \includegraphics[width=1.0\columnwidth]{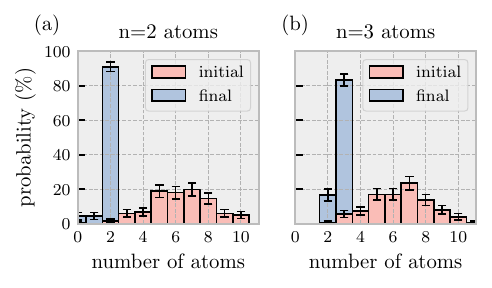}
    \caption{A histogram to visualize the results of the tweezer-rearrangement process. The numbers of probabilistically loaded tweezer-array traps are shown in light red and the numbers of atoms after the rearrangement process are shown in light blue. (a) An ordered array with a target size of two atoms is realized. (b) The target size of the array is three atoms.
    }
    \label{fig:fig4}
\end{figure}

We sort atoms using the \textit{dynamic} tweezer to prepare ordered arrays \cite{kim2016, barredo2016, endres2016} (Fig.~\ref{fig:fig3}). After imaging the stochastically loaded atomic pattern [Fig.~\hyperref[fig:fig3]{5(a)}], a computer calculates an optimal sorting sequence. Atoms are moved by overlapping the \textit{dynamic} tweezer with the \textit{static} ones, while ramping up the trap depth of the \textit{dynamic} tweezer to $U_\textit{dynamic}\sim 3\times U_\textit{static}$ within \qty{10}{\milli \second}. The atoms are then transported with a speed of \qty{5}{\micro \meter \per \milli \second} along a 2D trajectory in the x-y plane, optimized to avoid crossing other tweezers. Afterward, the atoms are released into the target \textit{static} tweezer trap by reducing $U_\textit{dynamic}$ to zero [Fig.~\hyperref[fig:fig3]{5(b)}]. Multiple movements can occur within the acquisition time of one picture of the EMCCD camera (\qty{350}{\milli \second}). If the atom rearrangement does not succeed, the sorting procedure is repeated until it succeeds. As soon as all target traps are filled [Fig.~\hyperref[fig:fig3]{5(c)}], redundant atoms are released from the tweezer-array traps by moving and releasing them in free space. In Fig.~\ref{fig:fig4}, we report an example of our results. We use an $8\times3$ tweezer array and load on average approximately 6.4 atoms. We then target ordered arrays of two and three atoms upon sorting, achieving a success probability of \qty{88}{\percent} and  \qty{82}{\percent}, respectively. These probabilities include the finite number of atoms initially transferred into the cavity and the tweezer capture probability, as well as the lifetime of the atoms in the tweezers, which is in the order of $\qty{30}{\second}$. 

\begin{figure}[t]
    \centering
    \includegraphics[width=1.0\columnwidth]{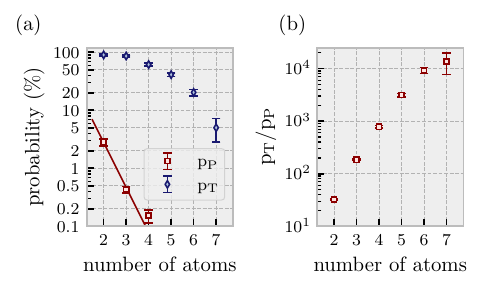}
    \caption{The atom-loading probability with and without the tweezers. (a) In red, we show the loading probability of $N$ atoms with purely probabilistic loading $p_\text{P}$ \cite{nussmann2005}. The blue data points show the loading probability $p_\text{T}$ employing the \textit{dynamic} tweezer to rearrange the atoms loaded into the \textit{static} tweezer array. 
    (b) The ratio of the loading probabilities employing the tweezers and the probabilistic loading $p_\text{T}/p_\text{P}$. This shows a tremendous increase in the atom-loading probability. The limited number of stochastically filled tweezers leads to a flattening of this curve for seven atoms. This can be overcome by increasing the MOT size or the number of tweezers.}
    \label{fig:atomLoading}
\end{figure}

Optical tweezers significantly enhance the atom-loading efficiency compared to previous probabilistic approaches. The success probability $p_\text{P}$ of loading and positioning atoms within the cavity has been merely 3\% for two atoms \cite{welte2018}, dropping exponentially to below 0.5\% and 0.15\% for three and four atoms, respectively [Fig.~\hyperref[fig:atomLoading]{7(a)}]. This severe scalability limitation is now overcome by the tweezer-based method, which allows for an improvement of more than an order of magnitude in the loading probability $p_\text{T}$ for two atoms and over 4 orders of magnitude for configurations with six and seven atoms [see Fig.~\hyperref[fig:atomLoading]{7(b)}].  Crucially, $p_\text{T}$ does not exhibit the exponential drop-off with increasing atom numbers. The observed decrease is mostly due to the limited number of atoms initially loaded, which can be increased in the future by using a larger intracavity tweezer array and a higher rubidium background gas pressure for a larger MOT.

\subsection{Controlled transfers to an optical lattice}
\label{subsec:Controlled_transfers}

\begin{figure}[t]
    \centering
    \includegraphics[width=1.0\columnwidth]{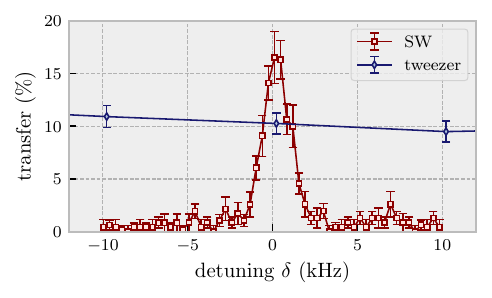}
    \caption{The Raman spectrum measured with the atom trapped in the \qty{1064}{\nano \metre} standing-wave trap (SW) and in the tweezer. Fluctuating vector light shifts caused by the oscillation of the atom inside the tightly focused optical tweezer broaden the spectrum. This hinders the coherent atom control and causes decoherence for magnetically sensitive states.}
    \label{fig:RamanSpectrum}
\end{figure}

The final step is the controlled transfer of the atoms to a 2D optical lattice, formed by a \qty{1064}{\nano \meter} standing-wave trap and the \qty{771}{\nano \meter} intracavity field. While optional, this step offers advantages for specific applications. Tightly focused optical beams such as tweezers generate extraordinary circular-polarization components near the focal point \cite{rosenfeld2008,thompson2013}. These components induce a Zeeman-state-dependent vector light shift that fluctuates while the atom is moving in the trap. To showcase this, we have pumped the atom into the $\ket{5S_{1/2}, F=1}$ manifold and we drive a two-photon Raman transition to transfer a part of the population into the $\ket{5S_{1/2}, F=2, m_\text{F}=2}$ state. We make this transition addressable in the frequency domain by applying a magnetic guiding field along the cavity axis, which causes a Zeeman splitting of \qty{100}{\kilo \hertz} \cite{hartung2024}. By measuring the population in $\ket{5S_{1/2}, F=2}$ while scanning the two-photon detuning $\delta$, we extract the Raman transition spectrum when the atom is trapped in the tweezer and when it is trapped in the standing-wave trap (see Fig.~\ref{fig:RamanSpectrum}). In the former, the spectrum is broadened by the fluctuating vector light shifts. Together with the additional decoherence induced by tweezer light scattering, this impedes high-quality coherent-qubit control, especially for qubits encoded in magnetic-field-sensitive states ($m_{\text{F}} \neq 0$). In contrast, standing-wave traps allow for high trap frequencies at beam waists much larger than the wavelength, effectively suppressing near-focal extraordinary polarization components. Furthermore, when the counterpropagating standing-wave beams are perfectly aligned, these undesired components are entirely eliminated through destructive interference between inward- and outward-propagating waves. Standing-wave traps, thus, inherently avoid vector light shifts and enable higher trap frequencies at lower laser power: the Raman spectrum in the latter case  shows a narrow feature in the kilohertz range, indicating a millisecond-long coherence time. 

\begin{figure}[t]
    \centering
    \includegraphics[width=1.0\columnwidth]{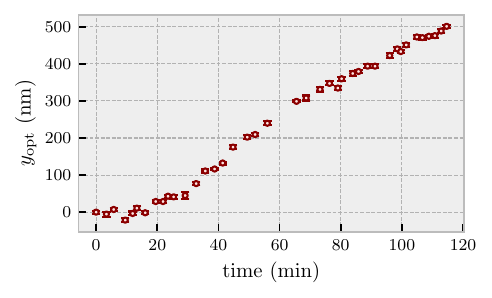}
    \caption{The thermal drifts of the position of the tweezers relative to the cavity. Due to light-induced heating of the cavity, the position of the strongest coupling of the cavity to the atom in the tweezer beam $y_{\text{opt}}$ changes on a time scale of minutes.}
    \label{fig:CavityDrift}
\end{figure}

An alternative solution is to use longer tweezer wavelengths, which reduces both photon scattering and the vectorial light shift but requires higher laser power. Larger power causes larger thermal fluctuations of the optical setup, which results in a slow drift between the tweezer position and the cavity, reducing the effective atom-cavity coupling strength. We have measured these drifts by repeating the measurement illustrated in Fig.~\ref{fig:pumpingAlongCavity} and recording the position of the maximum atom-cavity coupling $y_{\text{opt}}$. As shown in Fig.~\ref{fig:CavityDrift}, we observe a drift of the order of $\lambda_{\text{C}}/2$ in roughly $\qty{100}{minutes}$. Compared to the tweezer scheme, the use of the standing-wave traps completely solves this problem. By carefully selecting the intracavity trap wavelength (\qty{771}{\nano \meter} in our case), we ensure that atoms are consistently trapped near the maximum of the atom-cavity coupling \cite{reiserer2013}. Consequently, the atoms will move correspondingly with any thermal drift of the cavity, maintaining optimal positioning throughout the experiment.

The interatomic distance within the optical lattice is determined by the spacing of the tweezers in step (\hyperref[subsec:Stochastic_loading]{A}). This separation can be tuned by changing the rf signal applied to the AODs. Using the tweezer array of Fig.~\ref{fig:fig3}(c), the atoms are on average ten lattice sites of the \qty{1064}{\nano \metre} standing-wave trap apart.

Finally, we have measured the temperature of the atoms inside the optical lattice by performing Raman sideband spectroscopy \cite{reiserer2013}, i.e., by driving motional sidebands between the $\ket{5S_{1/2}, F=1}$ and $\ket{5S_{1/2}, F=2}$ manifold. We see that polarization gradient cooling brings the atoms close to the motional ground state, with an average occupation number of $\overline{n}_\text{x}=0.22, \, \overline{n}_\text{y}=0.35$.

\section{Conclusions and outlook}
\label{sec:Conclusion}

In this study, we have detailed our experimental techniques to prepare defect-free arrays of \isotope[87]{Rb} atoms within a microscopic high-finesse optical cavity using optical tweezers. We have shown an atom-positioning precision well below the cavity-field wavelength, which is essential for achieving maximal atom-cavity coupling. Compared to the probabilistic scheme employed previously \cite{nussmann2005}, the tweezer-based preparation scheme demonstrates an improvement of more than 4 orders of magnitude for preparing arrays of up to seven atoms, where the size of the array is currently limited by the initial number of atoms transferred into the cavity. Finally, we have demonstrated a combination of tweezers and optical lattices that integrates the flexibility of reconfigurable tweezers with the advantages of optical lattices. These advantages include strong confinement at moderate laser power, enhanced stability against thermal drift, and extended qubit coherence time.

Our results represent a key enabling step for building multiqubit quantum network nodes. Together with single-atom addressability and the efficient generation of atom-photon entanglement \cite{hartung2024}, we expect that these nodes can be used to implement multiplexed quantum repeater protocols \cite{sangouard2011} as well as error correction \cite{roffe2019} and entanglement purification \cite{kalb2017}. The integration of these capabilities will lead to functional large-scale quantum networks, supporting error-free long-distance communication. Future development can include implementing Rydberg-based atom-atom gates \cite{levine2019} within the cavity-coupled atomic array. This combination of network capability and local quantum information processing will facilitate the realization of distributed quantum computing machines \cite{Jiang2007,covey2023}, that can effectively scale up existing neutral-atom quantum computers.

Furthermore, the cavity-mediated interaction between individually positioned and controlled atomic qubits enables us to scale the two-qubit gate \cite{welte2018} to a multiqubit gate realized in a single shot \cite{duan2005, lin2006} and to generate large-scale atomic quantum states \cite{haas2014} with single-qubit control. Moreover, a larger atomic qubit array together with the cavity-enhanced atom-photon interface allows us to extend the recently demonstrated multiphoton entangled-state generation \cite{schwartz2016, thomas2022, thomas2024} to complex graph states of many photons with a programmable topology, which serve as a resource for photonic quantum computing \cite{Raussendorf2001} and communication \cite{borregaard2020}. Finally, the demonstrated subwavelength precision in atom positioning will enable dynamic control of the atom-cavity interaction strength for each atom individually. This precise control could enable fast cavity-enhanced midcircuit measurements of atomic qubits \cite{deist2022_2}, a key component in implementing error-correction quantum algorithms \cite{roffe2019}.

\section*{Acknowledgements}

We thank Maya Büki, Franz von Silva-Tarouca, Tobias Urban, Florian Furchtsam, Johannes Siegl, and Thomas Wiesmeier for experimental assistance. Additionally, we thank Stephan D\"{u}rr and Johannes Zeiher for fruitful discussions. This work was supported by the Bundesministerium f\"{u}r Bildung und Forschung through the project QR.X (16KISQ019), by the Deutsche Forschungsgemeinschaft under Germany’s Excellence Strategy (MCQST, Project No. 390814868), and by the European Union’s Horizon Europe research and innovation program via the project QIA-Phase 1 (Grant Agreement No. 101102140). E.D. acknowledges financial support from the Max Planck-Harvard Research Center for Quantum Optics postdoctoral scholarship and from the ``laCaixa'' Foundation through the Junior Leader fellowship ID100010434, code LCF/BQ/PI23/11970036. S.W. acknowledges support from the Center for Integrated Quantum Science and Technology (IQST) and financial support from the Swiss National Science Foundation (SNSF) Postdoctoral Fellowship (Project No. TMPFP2$\_$210584) and the Carl-Zeiss-Stiftung Center for Quantum Photonics.\\

All authors contributed to the experiment, the analysis of the results, and the writing of the manuscript.

\section*{Data Availability}

The data underlying the figures are deposited at Zenodo
\cite{seubert2025}. All other data needed to evaluate the conclusions in
the paper are present in the main text.

\section*{Appendix}
\label{sec:Appendix}
\renewcommand\thesubsection{\arabic{subsection}}
\subsection{Tweezer trap frequencies}
\label{subsec:Tweezer_trap}

\begin{figure}[t]
    \centering
    \includegraphics[width=1.0\columnwidth]{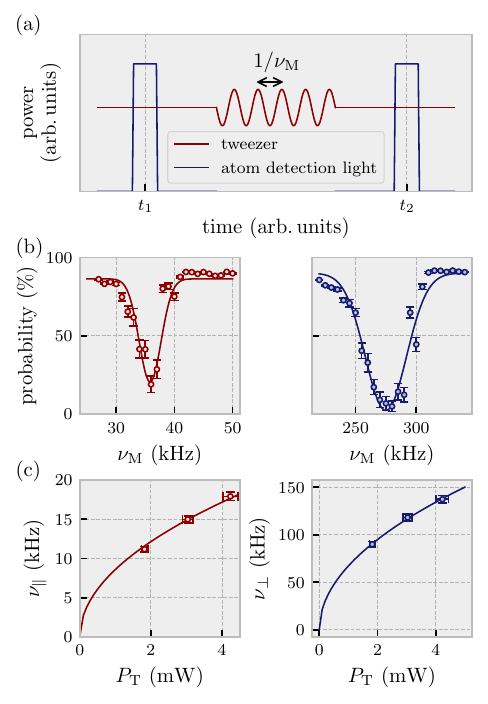}
    \caption{Characterization of the tweezer trap frequencies.     (a) The experimental sequence. The presence of the atom is checked at time $t_1$, followed by an amplitude modulation of the tweezer power at a frequency $\nu_\text{M}$ and a subsequent check of the atom presence at time $t_2$. (b) The atomic survival probability $p(\nu_\text{M})$ as a function of the modulation frequency $\nu_\text{M}$ for a tweezer power of \qty{4.2}{\milli \watt} at $\lambda_\text{T} = \qty{800.12}{\nano \metre}$. We find two frequencies at which $P(\nu_\text{M})$ drops, corresponding to $2\nu_{||}=2\times \qty{17.91(5)}{\kilo \hertz}$ and $2\nu_\perp = 2\times \qty{137.5(5)}{\kilo \hertz}$. (c) By measuring the perpendicular and parallel trap frequency for different optical powers, we determine the beam waist $w_0 = \qty{1.414(4)}{\micro \metre}$ and the Rayleigh range $z_\text{R} = \qty{7.8(1)}{\micro \metre}$ using Eq. \hyperref[eq:WaistVsTrapFrequency]{(A1)}.}
    \label{fig:T_TrapFrequency}

\end{figure}

We characterize the tweezer transversal $\nu_\perp$ (in the x-y plane) and longitudinal $\nu_{||}$ (along the $z$ axis) trap frequencies by trapping an atom in a single tweezer and modulating the tweezer amplitude at a frequency $\nu_\text{M}$ for a short time (Fig.~\ref{fig:T_TrapFrequency}). When $\nu_\text{M}=2\nu_\perp$ or  $\nu_\text{M}=2\nu_{||}$, the atom is parametrically heated and rapidly expelled from the tweezer trap. By measuring the atomic survival probability as a function of the modulation frequency $p(\nu_\text{M})$, it is thus possible to identify the trap frequencies. To measure $p(\nu_\text{M})$, we use cavity-enhanced fluorescence to check the presence of an atom in the tweezer. To this end, we lock the cavity resonant to the $\ket{5S_{1/2}, F=2, m_\text{F}=2}\rightarrow \ket{5P_{3/2}, F=3, m_\text{F}=3}$ transition and illuminate the atom through the $45^{\circ}$ axis with cavity-resonant light and a repumper. The latter pumps the atoms into the $\ket{5S_{1/2}, F=2}$ state.  Together with the cavity, the former stimulates a large atomic Purcell-enhanced scattering in the cavity mode that allows us to reveal the presence of an atom by monitoring the cavity output using a single-photon detector.  To measure $p(\nu_\text{M})$, we first check the presence of the atom at time $t_1$, then modulate the tweezer amplitude at a frequency $\nu_\text{M}$ for \qty{1}{\milli \second}, and finally check for the atom again at time $t_2$ [Fig.~\hyperref[fig:T_TrapFrequency]{10(a)}]. $p(\nu_\text{M})$ is given by the conditional probability of the atom being present at time $t_2$, given its presence at time $t_1$. Our results show two drops of $p(\nu_\text{M})$ [Fig.~\hyperref[fig:T_TrapFrequency]{10(b)}] which identify the two characteristic trap frequencies $\nu_{||}= \qty{17.91(5)}{\kilo \Hz}$ and $\nu_\perp = \qty{137.5(5)}{\kilo \Hz}$, corresponding to a harmonic oscillator length ($\sqrt{\hbar/(m_{^{87}\text{Rb}} \, \omega_\text{trap})}$) of $\qty{80.6(1)}{\nano \meter}$ and $\qty{29.08(5)}{\nano \meter}$. These data have been taken with a tweezer wavelength $\lambda_\text{T}=\qty{800.12}{\nano \meter}$ and a power of $P_{\text{T}} = \qty{4.2}{\milli \watt}$. The tweezer waist $w_0$ and Rayleigh range $z_{\text{R}}$ can be extracted by scanning the laser power, measuring the trap frequencies, and fitting the data using the relation \cite{grimm2000, sortais2007}
\begin{equation}
\tag{A1}
\label{eq:WaistVsTrapFrequency}
\begin{split}
&(2 \pi \times \nu_{\perp})^2 = \frac{4 \hat{U}}{m w_0^2} \\
&(2 \pi \times \nu_{\parallel})^2 = \frac{2 \hat{U}}{m z_{\text{R}}^2} \;\\ \text{with} \; &\hat{U} = \frac{\lambda_{\text{T}}^3 \Gamma}{16 \pi^2 c} \left(\frac{1}{\Delta_1} + \frac{2}{\Delta_2}\right) \frac{2 P_{\text{T}}}{\pi w_0^2}.
\end{split}
\end{equation}
Here, $m$ is the mass of \isotope[87]{Rb} atoms, $\Gamma$ is the linewidth of the $D_2$ line, and $\Delta_1$ and $\Delta_2$ are the detuning between the wavelength of the optical tweezers and the $D_1$ and $D_2$ lines, respectively. We obtain $w_0 = \qty{1.414(4)}{\micro \meter}$ and $z_{\text{R}} = \qty{7.8(1)}{\micro \meter}$.

\subsection{Superresolving tweezer imaging using a single atom}
\label{subsec:Super-resolving}

\begin{figure}[t]
    \centering
    \includegraphics[width=1.0\columnwidth]{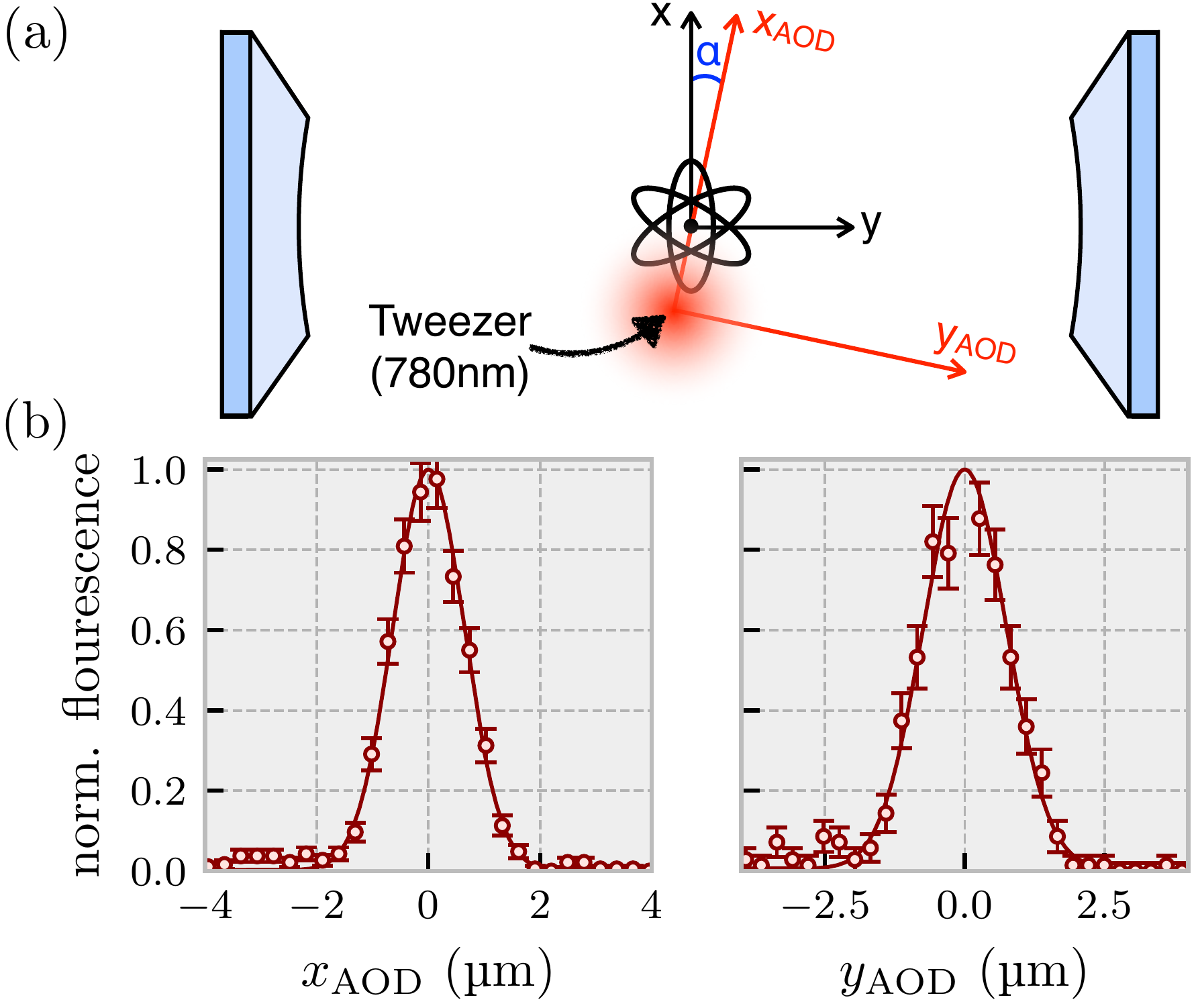}
    \caption{The subwavelength spatial mode imaging of the tweezer beam mode. (a) An atom-cavity resonant \qty{780}{\nano \metre} probe light is injected in the \textit{dynamic} tweezer path and its position is scanned on the cavity plane while one atom is trapped in the 2D lattice. The axes along which the beams move, $x_{\text{AOD}}$ and $y_{\text{AOD}}$,  are rotated by an angle $\alpha$ with respect to the cavity reference system $x$ and $y$. (b) The atomic fluorescence counts measured at the cavity output as a function of the probe position along $x_{AOD}$ (left) and $y_{AOD}$ (right). The fluorescence is directly proportional to the probe intensity, from which we can reconstruct the probe mode shape. We find two Gaussian-shape waists, $w_{\text{x}_{\text{AOD}}} = \qty{1.28(11)}{\micro \metre}$ and $w_{\text{y}_{\text{AOD}}} = \qty{1.49(13)}{\micro \metre}$, along $x_{\text{AOD}}$ and $y_{\text{AOD}}$, respectively. }
    \label{fig:Rotation_and_scan}
\end{figure}

Inferring the tweezer waist $w_{0}$ and the Rayleigh range $z_\text{R}$ by measuring the tweezer trap frequencies gives only very moderate information on the tweezer wave front and shape. For a more in-depth analysis, we have developed a technique that employs a single atom as a point probe to image the tweezer mode. For this, we inject \qty{780}{\nano \meter} probe light (resonant with the cavity and the $\ket{5S_{1/2}, F=2, m_\text{F}=2}\rightarrow \ket{5P_{3/2}, F=3, m_\text{F}=3}$ transition) into the same spatial mode as the \textit{dynamic} tweezers, i.e., using the same optical fiber. Note that despite a slight mismatch between the probe and tweezer wavelengths ($\qtyrange{797}{807}{\nano \metre}$), the probe is an effective indicator for detecting optical distortions at both wavelengths. We then load one atom at the cavity center using the optical lattice and scan the probe light position by tuning the 2D-AOD frequency. As soon as the probe light illuminates the atom, the cavity stimulates a strong atomic fluorescence emission in the cavity mode. By monitoring the photon flux as a function of the probe position, we can effectively image the spatial light mode with a resolution that is only limited by the precision with which we convert the AOD radio frequency to position ($\{\overline{\sigma}_\text{x}, \overline{\sigma}_\text{y}\} = \{\qty{47(5)}{\nano \metre}, \qty{46(6)}{\nano \metre}\}$: see Sec. \hyperref[sec:Subwavelength_positioning]{III}), which is well below the imaging wavelength. Our results are shown in Fig.~\ref{fig:Rotation_and_scan}. We note that the \textit{dynamic} tweezer axes are rotated by an angle of $\alpha = \ang{14.03(5)}$ with respect to the cavity axis [see Fig.~\hyperref[fig:Rotation_and_scan]{11(a)}]. Note that this angle is slightly different from the values given in Sec. \hyperref[sec:Subwavelength_positioning]{III} because of the different wavelength (\qty{780}{\nano \meter} instead of \qty{807}{\nano \meter}). In Fig.~\hyperref[fig:Rotation_and_scan]{11(b)}, we show the atomic fluorescence at the cavity output measured with a single-photon detector when the probe light position is scanned along the rotated AOD axes, $x_{\text{AOD}}$ and $y_{\text{AOD}}$, respectively. Our data show that the tweezer mode has a Gaussian shape in both directions, with waists $w_{\text{x}_{\text{AOD}}} = \qty{1.28(11)}{\micro \metre}$ and $w_{\text{y}_{\text{AOD}}} = \qty{1.49(13)}{\micro \metre}$. The difference between $w_{\text{x}_{\text{AOD}}}$ and $w_{\text{y}_{\text{AOD}}}$ indicates a slight astigmatism of the beam which, however, does not significantly affect the performance of the system. Compared to standard techniques based on a camera with a zoom objective, our method allows imaging of the mode \textit{in situ} and does not rely on the precise alignment of additional imaging optics. Furthermore, this technique can be used to image the position of atoms on time scales of \qty{}{\micro \second}, much shorter than the exposure times of regular cameras.



\end{document}